\documentclass[10pt,a4paper,twocolumn,superscriptaddress,aps,prd,longbibliography,nofootinbib]{revtex4-2}
\usepackage[utf8]{inputenc}
\usepackage{lmodern}
\usepackage[T1]{fontenc}
\usepackage{amsmath}
\usepackage{amsfonts}
\usepackage{amssymb}
\usepackage{bbold}
\usepackage{graphicx}
\usepackage[dvipsnames]{xcolor}
\usepackage{float}
\usepackage{multirow}
\usepackage{soul}
\usepackage{physics}
\usepackage{verbatim} 
\usepackage{placeins} 
\usepackage[colorlinks, allcolors=violet]{hyperref}
\usepackage{acronym}
\usepackage{bm}
\usepackage[capitalize]{cleveref} 

\makeatletter
\AtBeginDocument
 {
   \def\ltx@label#1{\cref@label{#1}}
   \def\label@in@display@noarg#1{\cref@old@label@in@display{#1}}
\def\label@in@mmeasure@noarg#1{%
    \begingroup%
      \measuring@false%
      \cref@old@label@in@display{#1}
    \endgroup}%
 } %
\makeatother

\newcommand{\fett}[1]{\textit{\textbf{#1}}}

\usepackage{blindtext}

\def\ket#1{|#1\rangle}

\let\s\textsubscript
\let\vec\bm

\newacro{EL-PH}{electron-phonon}
\newacro{ARPES}{angle-resolved photoemission spectroscopy}
\newacro{BZ}{Brillouin zone}
\newacro{CDW}{charge-density wave}
\newacro{DFPT}{density-functional perturbation theory}
\newacro{DFT}{density-functional theory}
\newacro{ISD}{inverse Star of David}
\newacro{SD}{Star of David}
\newacro{SI}{Supplementary Information}
\newacro{VHS}{Van Hove singularity}
\newacro{cDFPT}{constrained density-functional perturbation theory}
\newacroplural{ISD}[SD]{inverse Stars of David}
\newacroplural{SD}[SD]{Stars of David}
\newacroplural{VHS}[VHS]{Van Hove singularities}

\begin{document}

\author{Stefan Enzner}
 \affiliation{Institut f\"ur Theoretische Physik und Astrophysik and W\"urzburg-Dresden Cluster of Excellence ct.qmat, Universit\"at W\"urzburg, 97074 W\"urzburg, Germany}

\author{Jan Berges}
 \affiliation{U Bremen Excellence Chair, Bremen Center for Computational Materials Science, and MAPEX Center for Materials and Processes, University of Bremen, 28359 Bremen, Germany}
 \author{Arne Schobert}
 \affiliation{Institut f\"ur Theoretische Physik, Universit\"at Bremen, 28359 Bremen, Germany}
 \affiliation{I. Institute of Theoretical Physics, University of Hamburg, 22607 Hamburg, Germany}
 \author{Dongjin Oh}
 \affiliation{Department of Physics, Massachusetts Institute of Technology, Cambridge, MA 02139, USA}
 \author{Mingu Kang}
 \affiliation{Department of Physics, Massachusetts Institute of Technology, Cambridge, MA 02139, USA}
  \author{Riccardo Comin}
 \affiliation{Department of Physics, Massachusetts Institute of Technology, Cambridge, MA 02139, USA}
 \author{Ronny Thomale}
 \affiliation{Institut f\"ur Theoretische Physik und Astrophysik and W\"urzburg-Dresden Cluster of Excellence ct.qmat, Universit\"at W\"urzburg, 97074 W\"urzburg, Germany}
\author{Tim Wehling}
 \affiliation{I. Institute of Theoretical Physics, University of Hamburg, 22607 Hamburg, Germany}
 \affiliation{The Hamburg Centre for Ultrafast Imaging, 22761 Hamburg, Germany}
 \author{Domenico Di Sante}
 \affiliation{Department of Physics and Astronomy, University of Bologna, 40127 Bologna, Italy}

\author{Giorgio Sangiovanni}
\email{Contact author: giorgio.sangiovanni@uni-wuerzburg.de}
 \affiliation{Institut f\"ur Theoretische Physik und Astrophysik and W\"urzburg-Dresden Cluster of Excellence ct.qmat, Universit\"at W\"urzburg, 97074 W\"urzburg, Germany}

\date{\today}


\title{Phonon Fluctuation Diagnostics: Origin of Charge Order in AV$_3$Sb$_5$ Kagome Metals}

\begin{abstract}
The microsopic origin of the \protect\acf{CDW} in AV\s3Sb\s5 (A = K, Rb, Cs) kagome metals remains a longstanding question, often revolving around electron-phonon coupling and purely electronic mechanisms involving Van Hove scenarios, nesting, and sublattice interference. To reveal the processes driving the \ac{CDW} transition, we combine \emph{ab-initio} calculations analysis of the phonon self-energy and \ac{ARPES}. 
Our momentum-resolved study, supported by \ac{ARPES} data, reveals that lattice instabilities in the V-135 family of kagome metals appear to also be driven by electronic states far from high-symmetry points, where these states exhibit the strongest coupling with the phonon modes responsible for the CDW distortion. Footing on an interpretation scheme based on phonon fluctuation diagnostics, our work challenges and revises theories that so far have exclusively attributed \ac{CDW} formation to nesting effects close to the Fermi level.    
\end{abstract}
\maketitle

\textit{Introduction --} Kagome metals have been an emerging ground for exotic states of electronic matter \cite{neupert2022review,Wilson2024, Wang2023}. Instabilities in kagome metals are hypothesized to originate in simplified theoretical models from an intricate interplay of topology, electron-electron interactions, and enhanced availability of states due to \ac{VHS} at the Fermi level \cite{Neupert, Comin, proPhononK, Kiesel1, Kiesel2,scammel2023exciton,ingham2025vestigialorderexcitonicmother}. Yet, kagome metals are real materials featuring higher degree of complexity than what is encoded in few-bands tight-binding fermionic-only models, so that the identification of electronic degrees of freedom relevant for the various instabilities remains a major outstanding point in the field. Here, we identify the electronic degrees of freedom responsible for the lattice instabilities in kagome metals directly. We find that these instabilities, across a series of compounds, involve electronic states deep inside the \ac{BZ}, i.e., away from high symmetry points. This \emph{ab-initio} based finding challenges current understanding based on low-energy \ac{VHS} scenarios, nesting, and sublattice interference for Fermi surface instabilities \cite{Neupert, Comin, proPhononK, Kiesel1, Kiesel2, Wang13, Profe}. Capable to adequately tracking down electronic degrees of freedom far of the Fermi level, our findings shed new light on the understanding of recent \acf{ARPES}.

\begin{figure}
    \includegraphics[width=0.95\linewidth,trim={6.5cm 1.2cm 11.3cm 0.7cm},clip]{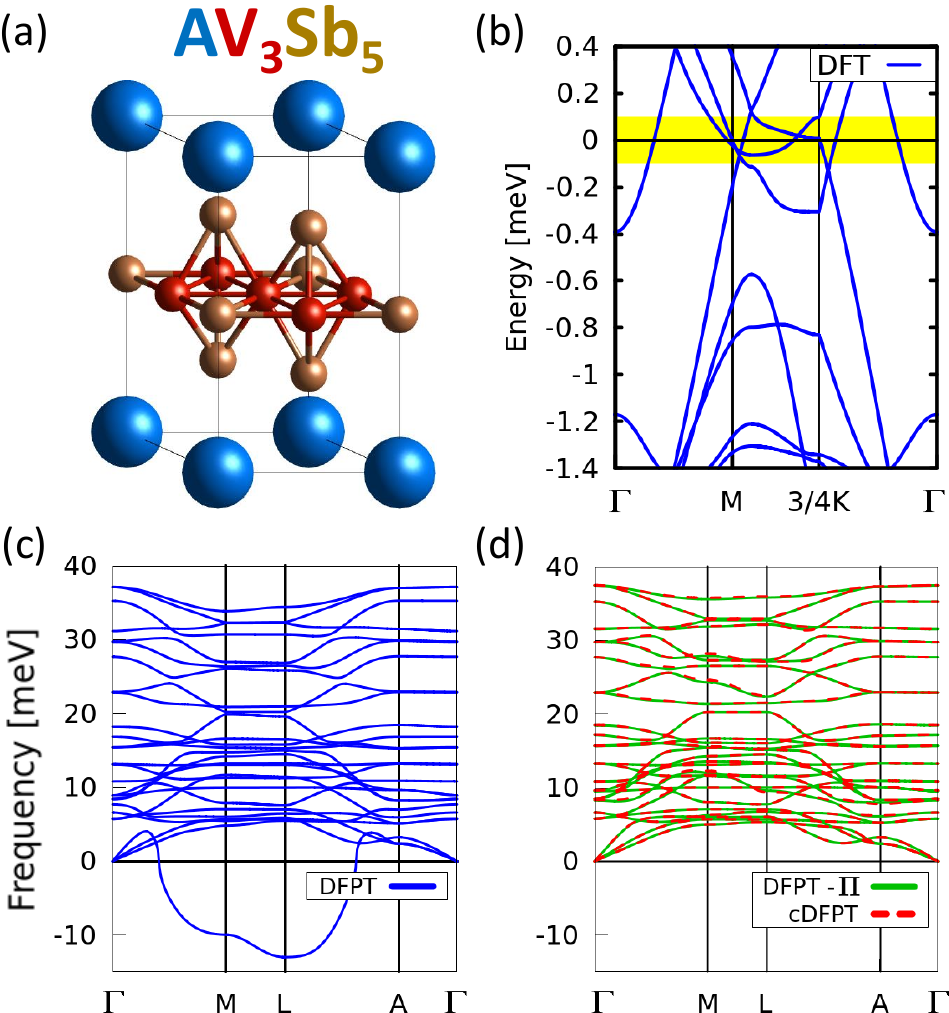}
   \caption{(a) Primitive unit cell of AV\s3Sb\s5. (b) CsV\s3Sb\s5 band structure along path shown in \cref{fig:self_energy}. The highlighted region indicates the active subspace  $\mathcal A$ used for the \ac{cDFPT} calculation. (c) Phonon dispersion of CsV\s3Sb\s5 with instability mode at $\vec q = \mathrm M$ and $\vec q = \mathrm L$ obtained from DFPT. (d) Phonon dispersion from cDFPT and ``unscreening'' the DFPT result with the approximate ($g^{\mathcal A} g^* \rightarrow |g|^2$) phonon self-energy.}
    \label{fig:instability}
\end{figure}

The family of AV\s3Sb\s5 (A = K, Rb, Cs) kagome superconductors has been heavily investigated due to its interesting lattice, frustration, topology, quantum geometry, chiral ordering, time-reversal symmetry breaking, flat bands, and the occurrence of a\,\acf{CDW} \cite{Ortiz_sc1, Ortiz_sc2, Transition_K_Hasan, Jiang_2023}. In particular, the origin of the charge ordering instability and its connection to superconductivity are currently under intense discussion \cite{Transition_Cs}. 
The members of the AV\s3Sb\s5 group show a \ac{CDW} transition below 78\,K--103\,K, where the structure undergoes a $2 \times 2 \times 2 (4)$ reconstruction \cite{Transition_K2, Transition_K1, Transition_Rb, Transition_Cs}. 
At the level of a single layer (2D), the compound is assumed to adopt a \ac{SD} or an \ac{ISD} configuration. Experimentally, the 135 V-based kagome exhibit different 3D ground-state reconstructions, where \ac{SD} and or \ac{ISD} can be stacked with or without a $\pi$-shift \cite{Comin222, Ortiz2021}.
The rearrangement can be directly linked to the theoretically observed phonon instabilities around the M- and L-points. We will put a particular focus on $\vec q = \mathrm L$; this is because we find it to exhibit the largest imaginary phonon frequency (\cref{fig:instability}(c)), even though either combinations of M and L can in principle contribute to the $2 \times 2 \times 2$ reconstruction (see \ac{SI} for $\vec q = \mathrm M$ analysis \cite{supp}).

The structure of V-based 135 kagome compounds is built up by a vanadium kagome lattice (red \cref{fig:instability}(a)), surrounded by Sb atoms from top and bottom as well as inside the kagome hexagon. These V\s3Sb\s5 layers are separated by the  A-atoms and determine the V~\textit{d} and Sb~\textit{p} dominant orbital contribution to the electronic structure around the Fermi level.
It is this multi-orbital behavior complicating the understanding of the low-energy physics, and rendering a simple one-orbital kagome description insufficient \cite{Neupert, Ronny_PRL}.

Instead of building toy models with specific orbital character to investigate the lattice instabilities, we pinpoint the electronic processes and the related electronic states based on an \emph{ab-initio} approach derived from \ac{DFT} and \ac{DFPT}. This analysis is performed by examining the $\vec k$-momentum-resolved phonon self-energy that can be directly connected to the microscopic origin of the \ac{CDW}, which we refer to as phonon fluctuation diagnostics \cite{Berges2020, elec_fluct}. We express the phonon self-energy as a product of the electron susceptibility and electron-phonon coupling, resolved in electronic momentum space. Therefore, phonon fluctuation diagnostics allows to discriminate electronic against electron-phonon contributions to a given lattice instability. In addition, our comparison of fluctuation diagnostics with ARPES measurements provides an unprecedented confirmation of our findings.

The hitherto common narrative with regard to the origin of the \ac{CDW} mechanism triangulates between the central themes of electronic interactions, phonon modes and electron-phonon coupling, as well as properties of Fermiology such as the nature and Fermi level distance of VHS, Fermi surface nesting, and sublattice interference \cite{Neupert, Comin, proPhononK, Kiesel1, Kiesel2, Wang13, Profe}. One scenario of choice derives the charge order instability from the large density of states of the \ac{VHS} at the M- and L-points close to the Fermi level. The phonon instability vectors $\vec q = \mathrm M$ and $\vec q = \mathrm L$ link these high-symmetry points, suggesting a Peierls mechanism resulting in a $2 \times 2$ instability \cite{Neupert}. This explanation suggests a correlation between charge order and VHS proximity to the Fermi level, which is challenged by kagome CsTi\s3Bi\s5. There, the energy of the \ac{VHS} can be tuned to the Fermi level with doping, yet no \ac{CDW} is observed \cite{CsTiBi}. Even more drastically, recent findings reveal an anticorrelation between the \ac{CDW} order parameter and the distance of the \ac{VHS} to the Fermi level. Setting aside subtleties in the ordering propensities of charge vs nematic order \cite{PhysRevB.110.245138} and its substantiation in the distinction between ordinary and higher order \ac{VHS}, it suggests that the electronic states at the \ac{VHS} might not be the single driving mechanism for the \ac{CDW} \cite{anticorr}. 
Another related scenario centers around more general (Fermi surface) nesting, where, instead of specific high-symmetry points, entire lines in the \ac{BZ} are connected by the instability vectors \cite{Comin}. As much as this rational idea could explain the origin of the \ac{CDW} for an idealistic system with perfect nesting, it struggles to account for real band structures under pressure and strain, where the Fermi surface nesting becomes significantly worse yet the charge order instability persists largely unaffected.
While beforementioned scenarios imply a leading role of  electronic correlation effects with an at best largely coarse-grained modelling of electron-phonon coupling, the charge order instability could also be determined by a strong or $\vec k$-selective electron-phonon coupling \cite{proPhononK}. 

In this article, we seek to shed new light on the origin of the \ac{CDW} instabilities in AV\s3Sb\s5 by taking the modelling of phononic impact on charge order to the next level from an \emph{ab-initio} perspective. We find the nature of electron-phonon coupling to constitute the key component to the driving mechanism of charge order, and reach suggestive agreement between theoretical modeling and \ac{ARPES} results.

\textit{Instabilities and phonon self-energies --} To investigate the origin of the charge density wave (CDW), we first establish the connection between the phonon self-energy $\Pi$ and the lattice instability. 
The interatomic force constants, and thus the phonons, are largely determined by the response of the electron density to ionic displacements.
In the following, we show that the most relevant contributions come from the low-energy electrons with energies close to the chemical potential.
We denote this \emph{active subspace} of low-energy electronic states as $\mathcal A$.
The contribution of processes within this subspace to the force constants reads
\begin{equation}
    \label{eq:Pi}
    \Pi_{\vec q \mu \nu}^{\mathcal A}
    = \frac 2 N
    \sum_{\vec k m n}^{\mathcal A}
    g_{\vec q \mu \vec k m n}^{\mathcal A}
    \frac
        {f(\varepsilon_{\vec k + \vec q m}) - f(\varepsilon_{\vec k n})}
        {\varepsilon_{\vec k + \vec q m} - \varepsilon_{\vec k n}}
    g_{\vec q \nu \vec k m n}^*,
\end{equation}
with the Kohn-Sham energies $\varepsilon$, the corresponding occupations $f$, the (partially) screened electron-phonon coupling $g^{(\mathcal A)}$, the electronic and phononic wave vectors $\vec k$ and $\vec q$ and band indices $m, n$ and $\mu, \nu$, and the number of wave vectors summed over $N$.
While the summation is constrained to active states, $\{\ket{\vec k m}, \ket{\vec k {+} \vec q m}\} \subset \mathcal A$, the partially screened coupling $g^{\mathcal A}$ \emph{excludes} screening from scattering between such states. We utilize an active window $\mathcal A$ of 0.1\,eV around the Fermi level (indicated in \cref{fig:instability}(b)) and compute $\varepsilon$, $g$, and $g^{\mathcal A}$ using \ac{DFT}, \ac{DFPT}~\cite{Baroni2001}, and \ac{cDFPT}~\cite{Nomura2015}.

\begin{figure}[t!]
 \includegraphics[width=1\linewidth,trim={13.8cm 0cm 2.2cm 1.5cm},clip]{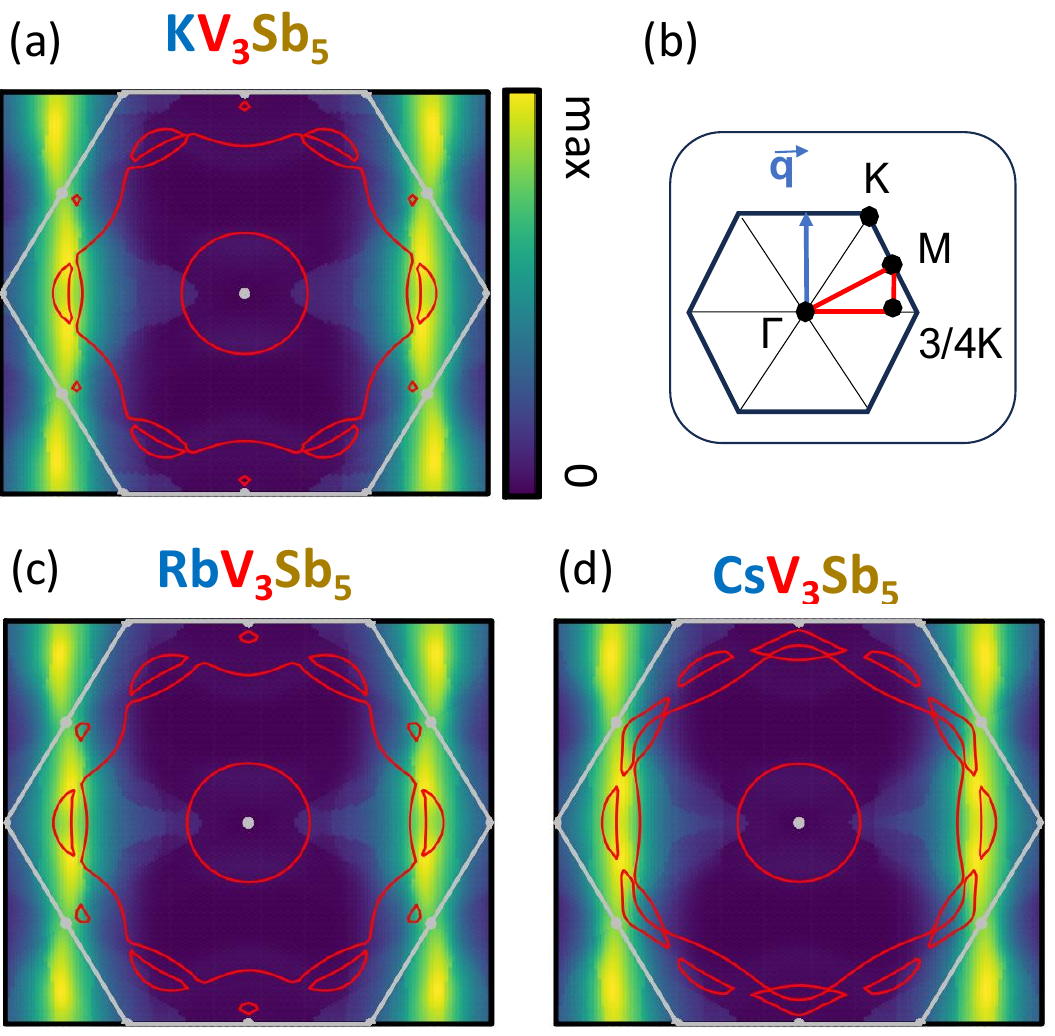}
    \caption{(a),(b) and (c) The $\vec k$-resolved electronic contributions to the phonon self-energy of AV\s3Sb\s5 with A = (K, Rb, Cs) for $\vec q = \mathrm L$, visualized in the $k_z = 0$ plane. The Fermi contour inside the electronic \ac{BZ} is shown in red. All members of the vanadium-135 kagome family exhibit similar self-energies with the main contributions occurring in the region between the M-points, however not directly at the M-point. The high-symmetry points and the in-plane component of the utilized $\vec q$ are shown in (b).}
    \label{fig:self_energy}
\end{figure}
Compared to the unstable \ac{DFPT} result in \cref{fig:instability}(c) that agrees well with the extensive literature, the \ac{cDFPT} approach with partially screening inside $\mathcal A$ stabilizes the structure (see \cref{fig:instability}(d)) \cite{Binghai, Consiglio2022}.
Similarly, by renormalizing the \ac{DFPT} frequencies with the phonon self-energy through 
\begin{equation}
    \Tilde{\omega}_{\vec q \mu \nu}^2 = \omega_{\vec q \nu}^2 \delta_{\mu \nu} + 2\sqrt{\omega_{\vec q \mu}\omega_{\vec q \nu}}\,\Pi_{\vec q \mu \nu},
    \label{eq_renorm}
\end{equation}
one can also avoid the phonon instabilities and reproducing the \ac{cDFPT} result very accurately  \cite{Berges2020}.
Here, $\Tilde{\omega}_{\vec q \mu \nu}$ and $\omega_{\vec q \nu}$ are the renormalized and \emph{ab-initio} phonon frequencies.
\begin{figure*}
    \includegraphics[width=0.8\linewidth,trim={0cm 2cm 0cm 7cm},clip]{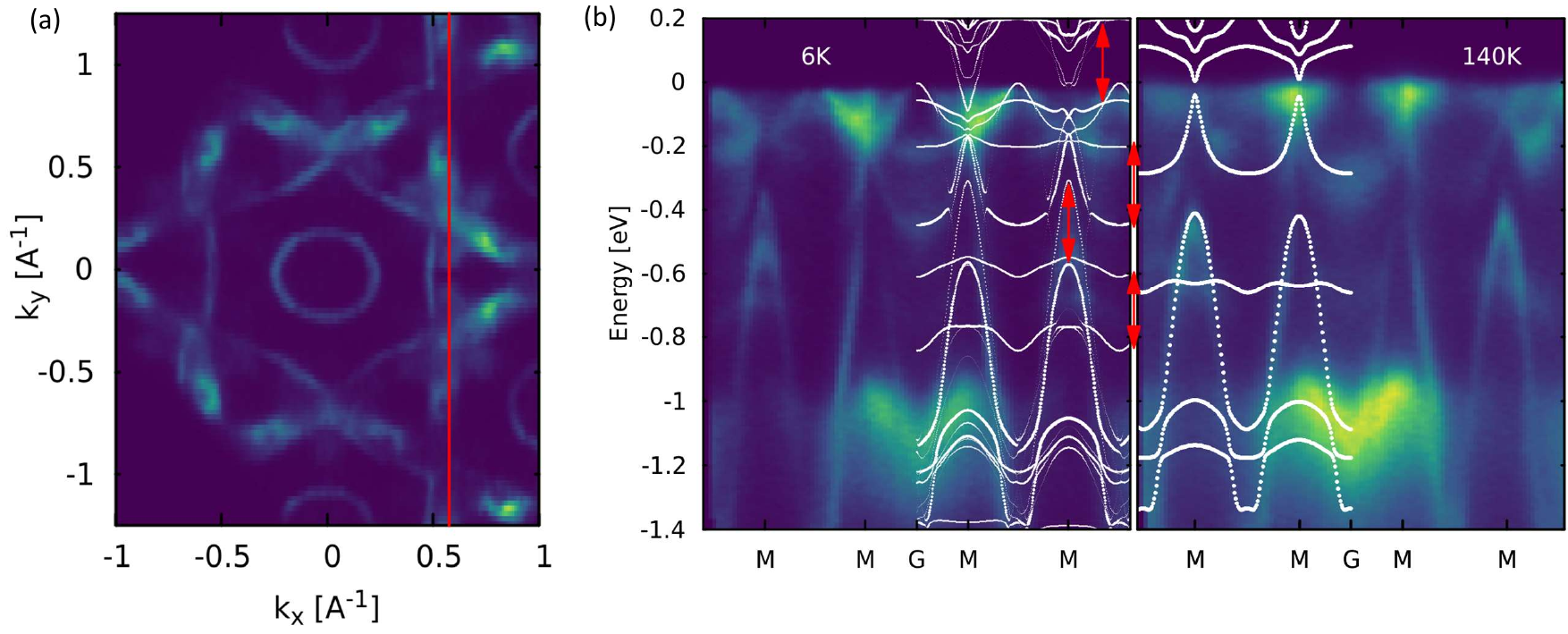}
    \caption{(a) Fermi surface of KV\s3Sb\s5 in the low-temperature phase with the path in red trough adjacent M-points. (b) \ac{ARPES} band structure below and above the \ac{CDW} temperature transition overlayed with \ac{DFT} bands the unfolded \ac{DFT} band structure of staggered \ac{ISD} $2 \times 2 \times 2$ phase and the bands of the primitive cell. Below the CDW transition the bands split up along the identified path between the M-points, as indicated with red arrows. 
    }
    \label{fig:bands}
\end{figure*}

The stabilization highlights that the CDW instability is encoded in the phonon self-energy or respectively the reduction in screening. The only phonon mode significantly altered by these approaches is the unstable one. This further emphasizes the direct connection of the phonon self-energy with the CDW instability.
Due to the agreement in \cref{fig:instability}(d), we approximate \cref{eq:Pi} with two screened $g$ for simplicity (as $g^{\mathcal A} \rightarrow g$ for $\mathcal A \rightarrow \emptyset$) \cite{Calandra2010, Berges2023}.

After establishing the critical role of the phonon self-energy regarding the \ac{CDW} instability in 135-V kagome structures, a comparison between the three members of the family is drawn. Based on \emph{ab-initio} results, the $\vec k$-resolved electronic contributions to the phonon self-energy at the wave vector $\vec q = \mathrm L$ of the instability is shown in \cref{fig:self_energy} (\cref{eq:Pi} before $\vec k$ summation). For clarity, we limit the analysis to one of three equivalent $\vec q = \mathrm L$ vectors, which causes the features to appear with only C\s2 symmetry. Nevertheless, with all three  $\vec q = \mathrm L$ the investigated properties retain the full C\s6 symmetry of the structure. Notably, the $\vec k$-dependence of the phonon self-energy is similar across all three materials, indicating a common origin and a consistent behavior of the \ac{CDW}. The highest amplitudes appear in the regions
between two adjacent M-points, which are connected by the in-plane component of the examined $\vec q$ vector associated with the instability. However, it is important to emphasize that this range does not extend fully to the M-point, where only minor phonon self-energy contributions are observed.
This observation deems the \ac{VHS} scenario unlikely as the dominant driver of the charge instability. Under this scenario, one would expect strongest phonon self-energy contributions near the M-points, where the density of states diverges.

\Cref{fig:self_energy} shows the Fermi surface of the three AV\s3Sb\s5 materials in red inside the electronic \ac{BZ}. Here the $\vec k$-resolved phonon self-energy is shown for \textit{\textbf{k}}$_\text{z}=0$, as we do not find a strong $k_\text{z}$ dependency (see \ac{SI} \cite{supp}).

The significance of the region between the M-points is further emphasized by comparing unfolded \ac{DFT} band structures in the energetically favored $2 \times 2 \times 2$ staggered \ac{ISD} supercell with \ac{ARPES} maps. This is done for KV\s3Sb\s5 as it was previously reported to exhibit the theoretically expected $2 \times 2 \times 2$ reconstruction at low temperatures \cite{Comin222, band_folding}. Both experimentally and theoretically, band splittings can be observed in \cref{fig:bands} along the indicated path through adjacent M-points as the temperature is reduced below the \ac{CDW} transition temperature. As indicated in \cref{fig:bands}(b), these splittings not only occur at the M-point but also between them, consistent with previous reports, further contradicting a dominant \ac{VHS} scenario \cite{Comin}. Additionally, the splitting is evident in the significant downshift of the intensity near the Fermi level around M.
By combining phonon fluctuation diagnostics with \ac{ARPES} we experimentally confirm the region in momentum space, which we theoretically identified from the analysis of the phonon self-energies role.

\textit{Phonon fluctuation diagnostics --}\label{ch:fluctuation}
\begin{figure*}
 \includegraphics[width=0.95\linewidth,trim={0cm 0cm 2.1cm 8cm},clip]{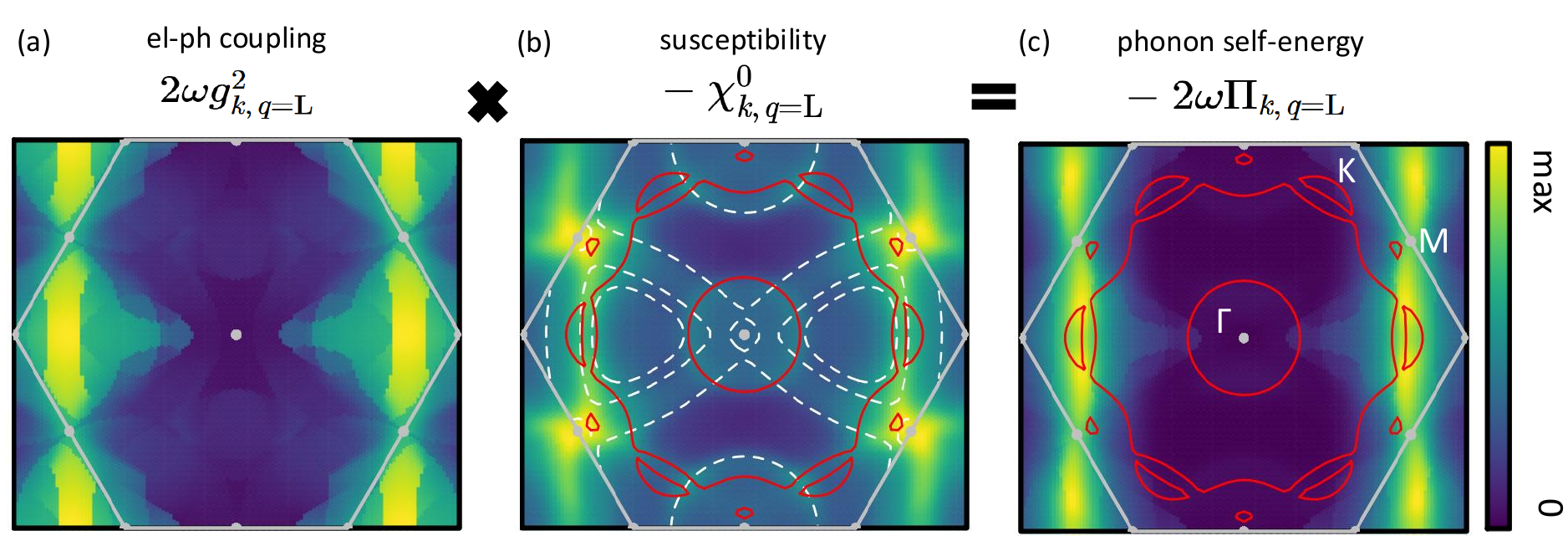}
    \caption{The $\vec k$-momentum-resolved phonon fluctuation diagnostics of the instability mode at $\vec q = \mathrm L$ for $k_z=0$. (a) The electronic susceptibility $\chi^0$, (b) the coupling matrix elements $2 \omega g^2$ (el-ph),  and (c) the phonon self-energy $2 \omega \varPi$ are shown. The solid red lines indicate the Fermi surface and the dashed lines Fermi surface shifted by $q = \mathrm L$ inside the electronic \ac{BZ}. The strong electron-phonon coupling between the $M$ points dominates the phonon self-energy.}
    \label{fig:fluc_diagnostic}
\end{figure*}
In the following, the phonon self-energy will be further investigated. To disentangle the electronic response and matrix elements, we consider the momentum-resolved phonon self-energy $\Pi_{\vec q \mu \nu \vec k m n}$ as a product of the electron-phonon coupling $g^2_{\vec q \mu \nu \vec k m n}$ and the electronic susceptibility $\chi^0_{\vec q \vec k m n}$.
This allows to distinguish the different contributions of the electronic density response to atomic displacements and enables us to identify the electronic processes behind the phonon softening associated with the origin of the \ac{CDW}. We representatively discuss KV\s3Sb\s5 and refer to the \ac{SI} for the other compounds \cite{supp}. The relevant formula are:
\begin{equation} \label{eq:product}
        \Pi_{\vec q \mu \nu \vec k m n} = g^2_{\vec q \mu \nu \vec k m n} \times \chi^0_{\vec q \vec k m n},
\end{equation}

\begin{equation}
     g^2_{\vec q \mu \nu \vec k m n} = g_{\vec q \mu \vec k m
n}^{(\mathcal A)} g_{\vec q \nu \vec k m n}^*.
\end{equation}

\begin{equation}\label{eq:sus}
    \chi^0_{\vec q \vec k m n} = \frac
        {f(\varepsilon_{\vec k + \vec q m}) - f(\varepsilon_{\vec k n})}
        {\varepsilon_{\vec k + \vec q m} - \varepsilon_{\vec k n}},
\end{equation}

We analyze the phonon mode at $\vec q = \mathrm L$ where the highest imaginary frequency can be found, which is indicative of the instability in the system (phonon branch indices $\nu = \mu = \text{soft mode}$). The corresponding \fett{k}-resolved electron-phonon coupling is obtained with \emph{ab-initio} calculations and depicted in \cref{fig:fluc_diagnostic}(a) for $k_z = 0$ (for computational details, see Methods).
We find the strongest electron-phonon coupling in the area between the M-points connected by the in-plane component of the chosen \fett{q} vector, whereas not at the M-points themselves. To illustrate the electron-phonon coupling coefficient relevant for the phonon self-energy, in \cref{fig:fluc_diagnostic}(a) we sum over the band indices $m$ and $n$ close to the Fermi level, as only these contribute to the product with the susceptibility (see \cref{eq:product}).
The electronic susceptibility evaluated based on \cref{eq:sus} from a Wannier model is shown in \cref{fig:fluc_diagnostic}(b) together with the $k_z=0$ Fermi surface in solid red lines. Additionally, a Fermi surface shifted by the instability wave vector $\vec q = \mathrm L$ is depicted with dashed lines to visualize possible low-energy transitions. As expected, the strongest contribution to the electron susceptibility can be found around the M-point (see \ac{SI} for band-resolved susceptibility \cite{supp}).

Nevertheless, the electron-phonon coupling only selects the specific \fett{k}-region between the M-points of the susceptibility for the contribution to the final phonon self-energy in \cref{fig:fluc_diagnostic}(c). This highlights the dominance of the electron-phonon matrix elements over electronic nesting or Van Hove effects in determining the phonon self-energy.
Such a behavior is a strong indicator for electron-phonon-coupling to be the driving factor of the \ac{CDW} instability.

In the case of a full \ac{VHS} or general nesting scenario, one would expect the susceptibility to dictate the phonon self-energy as to observe the main contributions around the M-point, where the significant nesting occurs. Such an origin of the instability is in contradiction to our result shown in \cref{fig:fluc_diagnostic}(c), where the electron-phonon coupling suppresses the susceptibility close to the M-point.

A similar behavior can be observed for various $k_z$ cuts and the RbV\s3Sb\s5 and CsV\s3Sb\s5 compounds (see \ac{SI} \cite{supp}). This suggests a common origin of the \ac{CDW} for the V-135 kagome group, even though different $2 \times 2 \times 2$ reconstructions are observed among these compounds, depending on temperature and pressure.

Our findings might crucially contribute to resolving the controversy on the origin of the \ac{CDW}, were different mechanisms have been proposed. Through disentangling electronic and electron-phonon coupling effects together with a $\vec k$-resolved analysis we are able to distinguish between the different scenarios explaining the CDW. Here, our findings suggest electron-phonon coupling as the dominant contribution to the origin of the \ac{CDW} in AV\s3Sb\s5 \cite{Roser, He2024, Xie22, Subires2023, GutierrezAmigo2024, proPhononK}, where electronically driven charge ordering propensity might exhibit a cooperative effect. Our identification of a specific $\vec k$-region away from high-symmetry points as a main manifestation regime for the lattice charge order instability is experimentally confirmed by \ac{ARPES} results.
This result leaves room for other many-body effects such as charge-bond order, loop-current order, pair-density waves and superconductivity to originate from the \ac{VHS} or M-point scattering without dominant phonon contributions. The ultimately complete picture of charge order in kagome metals will likely necessitate aspects of all microscopic sources for charge order formation, which is likely particularly true for the 135 family where electron correlation effects are assumed to be prominent. Our identification of phonon-mediated charge formation away from the BZ location of low energy density of states promises to carry over as a generic theme of all kagome metals. Recently, a crucial importance of phonons was observed in the ScV\s6Sn\s6 kagome metal, hinting at the generic significance of phonons in kagome systems \cite{Tuniz2023, Tuniz2, Bernevig}.

\textit{Conclusion --} Thanks to the combination of \emph{ab-initio} calculations and \ac{ARPES} measurements we largely resolve the long debated question about the origin of the \ac{CDW} in AV$_3$Sb$_5$ (A = K, Rb, Cs).
We utilized the idea of fluctuation diagnostics to investigate the interplay of lattice instabilities with electronic processes and electron-phonon coupling. By directly connecting the phonon self-energy to the lattice instability, we were able to disentangle the Fermiology and electron-phonon matrix elements effects. Here, we identify electron-phonon coupling as the driving effect for \ac{CDW} within all three V-135 materials.\\
From our momentum-resolved analysis of the response of the electron density to atomic displacements we find the dominant contributions away from high-symmetry points, and especially the M-point, where the kagome \ac{VHS} are located. We experimentally confirm that \ac{CDW} spectral signatures emerge between the M-points away from high-symmetry points. While this does not entirely rule out \ac{VHS} and ideal nesting scenarios as cause of the predicted and experimentally observed $2 \times 2 \times 2$ lattice instability, it stresses the unavoidable relevance of phonons in the eventual formation of high-temperature charge order in kagome metals.
It will be interesting to transfer our approach to other systems to understand existing or absent lattice instabilities. The kagome compounds with 135 stoichiometry offer an attractive platform for future investigations. This particularly applies to  CsTi$_3$Bi$_5$ not exhibiting a \ac{CDW} \cite{CsTiBi, CsTiBi2} and CsCr\s3Sb\s5 where different origins of the instability are currently discussed \cite{CsCrSb}.
{\noindent
}
\begin{scriptsize}
\end{scriptsize}


\textit{Acknowledgements --}
We gratefully acknowledge the Gauss Centre for Supercomputing e.V. (https://www.gauss-centre.eu) for funding this project by providing computing time on the GCS Supercomputer SuperMUC-NG at Leibniz Supercomputing Centre (https://www.lrz.de). We are grateful for funding support from the Deutsche Forschungsgemeinschaft (DFG, German Research Foundation) under Germany's Excellence Strategy through the W\"urzburg-Dresden Cluster of Excellence on Complexity and Topology in Quantum Matter ct.qmat (EXC 2147, Project ID 390858490), the University Allowance of the University of Bremen (EXC 2077, Project ID 390741603), FOR 5249 (QUAST, Project ID 449872909 (Project P3 and P5), EXC 2056 (Cluster of Excellence “CUI: Advanced Imaging of Matter”, Project No. 390715994) as well as through the Collaborative Research Center SFB 1170 ToCoTronics (Project ID 258499086).

\textit{Author contributions --}
S.E., J.B. and A.S. performed \emph{ab-initio} calculations and carried out the phonon fluctuation diagnostics analysis. D.O., M.K. and R.C. conducted the \ac{ARPES} experiments. T.W. and G.S. conceived and supervised this project. All authors contributed to the data interpretation and jointly wrote the manuscript.

\textit{Data availability --} 
The data that support the findings of this letter are openly available \cite{nomad, Zenodo}.




\nocite{QE1, QE2, wannier, EPW1, EPW2, PBE, pot, K_lattice, Cs_lattice, elphmod, vasp, vaspkit}
\bibliography{biblio}



\end{document}



\author{Stefan Enzner}
 \affiliation{Institut f\"ur Theoretische Physik und Astrophysik and W\"urzburg-Dresden Cluster of Excellence ct.qmat, Universit\"at W\"urzburg, 97074 W\"urzburg, Germany}

\author{Jan Berges}
 \affiliation{U Bremen Excellence Chair, Bremen Center for Computational Materials Science, and MAPEX Center for Materials and Processes, University of Bremen, 28359 Bremen, Germany}
 \author{Arne Schobert}
 \affiliation{Institut f\"ur Theoretische Physik, Universit\"at Bremen, 28359 Bremen, Germany}
 \affiliation{I. Institute of Theoretical Physics, University of Hamburg, 22607 Hamburg, Germany}
 \author{Dongjin Oh}
 \affiliation{Department of Physics, Massachusetts Institute of Technology, Cambridge, MA 02139, USA}
 \author{Mingu Kang}
 \affiliation{Department of Physics, Massachusetts Institute of Technology, Cambridge, MA 02139, USA}
  \author{Riccardo Comin}
 \affiliation{Department of Physics, Massachusetts Institute of Technology, Cambridge, MA 02139, USA}
 \author{Ronny Thomale}
 \affiliation{Institut f\"ur Theoretische Physik und Astrophysik and W\"urzburg-Dresden Cluster of Excellence ct.qmat, Universit\"at W\"urzburg, 97074 W\"urzburg, Germany}
\author{Tim Wehling}
 \affiliation{I. Institute of Theoretical Physics, University of Hamburg, 22607 Hamburg, Germany}
 \affiliation{The Hamburg Centre for Ultrafast Imaging, 22761 Hamburg, Germany}
 \author{Domenico Di Sante}
 \affiliation{Department of Physics and Astronomy, University of Bologna, 40127 Bologna, Italy}

\author{Giorgio Sangiovanni}
\email{e-mail: sangiovanni@physik.uni-wuerzburg.de}
 \affiliation{Institut f\"ur Theoretische Physik und Astrophysik and W\"urzburg-Dresden Cluster of Excellence ct.qmat, Universit\"at W\"urzburg, W\"urzburg, Germany}

\date{\today}


\title{Phonon Fluctuation Diagnostics: Origin of Charge Order in AV$_3$Sb$_5$ Kagome Metals: Supplementary Information}
\maketitle

\textit{Methods --} The \ac{ARPES} experiments were performed at Beamline 7.0.2 of the Advanced Light Source (MAESTRO). KV\s3Sb\s5 single crystals were cleaved inside an \ac{ARPES} chamber under ultra-high-vacuum conditions ($~3 \times10^{-11} \,\text{torr}$)
\newline

For our theoretical study of the kagome AV\s3Sb\s5 family with A = (K, Rb, Cs) we employed first-principles calculations based on \ac{DFT} and \ac{DFPT} theory as implemented in \textsc{Quantum ESPRESSO} \cite{QE1,QE2}. In order to obtain the electron-phonon couplings in a Wannier basis, the \textsc{Wannier90} software \cite{wannier} and the EPW code \cite{EPW1,EPW2} are utilized.
We apply the generalized gradient approximation \cite{PBE} and use norm-conserving pseudopotentials \cite{pot} with a plane-wave cut-off of 84 Ry.
We sampled the \ac{BZ} on a $12\times12\times8$ $\Gamma$ centered mesh and calculated the phonons on a $6\times6\times2$ grid. We employed a Gaussian smearing of 4\,mRy.
The lattice constants of the unit cells are chosen based on experimental parameters for 0\,GPa for K, Rb, and Cs respectively \cite{K_lattice, Transition_Rb, Cs_lattice} and the atomic positions relaxed until the forces are converged to below $10^{-7}$\,Ry/au.
To compare DFT band structure with experimental \ac{ARPES} data in Fig. 3(b) we utilize $k_z = 0.3$ DFT cuts. The $k_z$ dependence can be noticed in \cref{fig:kz_bands}. For the band structure calculations with the Vienna \textit{ab-initio} simulation package (\textsc{VASP}) \cite{vasp} and \textsc{VASPKIT} \cite{vaspkit} we used a plane wave energy cutoff of $270\ $eV.

For the \textsc{EPW} calculation, we create a 30-orbital Wannier Hamiltonian based on the V d-orbitals and the Sb p-orbitals. The phonon fluctuation diagnostics calculations were done with the \textsc{ELPHMOD} package \cite{elphmod}. Here we consider a temperature $k_\text{B}T \approx 5\,\text{meV}$ evaluated on a $100 \times 100$ $\mathbf{k}$ points grid. To visualize the electron-phonon coupling relevant for the bands around the Fermi level, we only consider the bands closest to the Fermi level as indicated in \cref{fig:sus_res}. The analysis is done for the $\mathbf{q}$ vectors M and L, where unstable phonons were observed. All color scales are normalized to the individual maximum of each quantity. \\
The input and relevant output files of the calculations are openly available \cite{nomad,zenodo}\\

\begin{figure*}
 \includegraphics[width=0.95\linewidth,trim={0cm 0cm 0cm 0cm},clip]{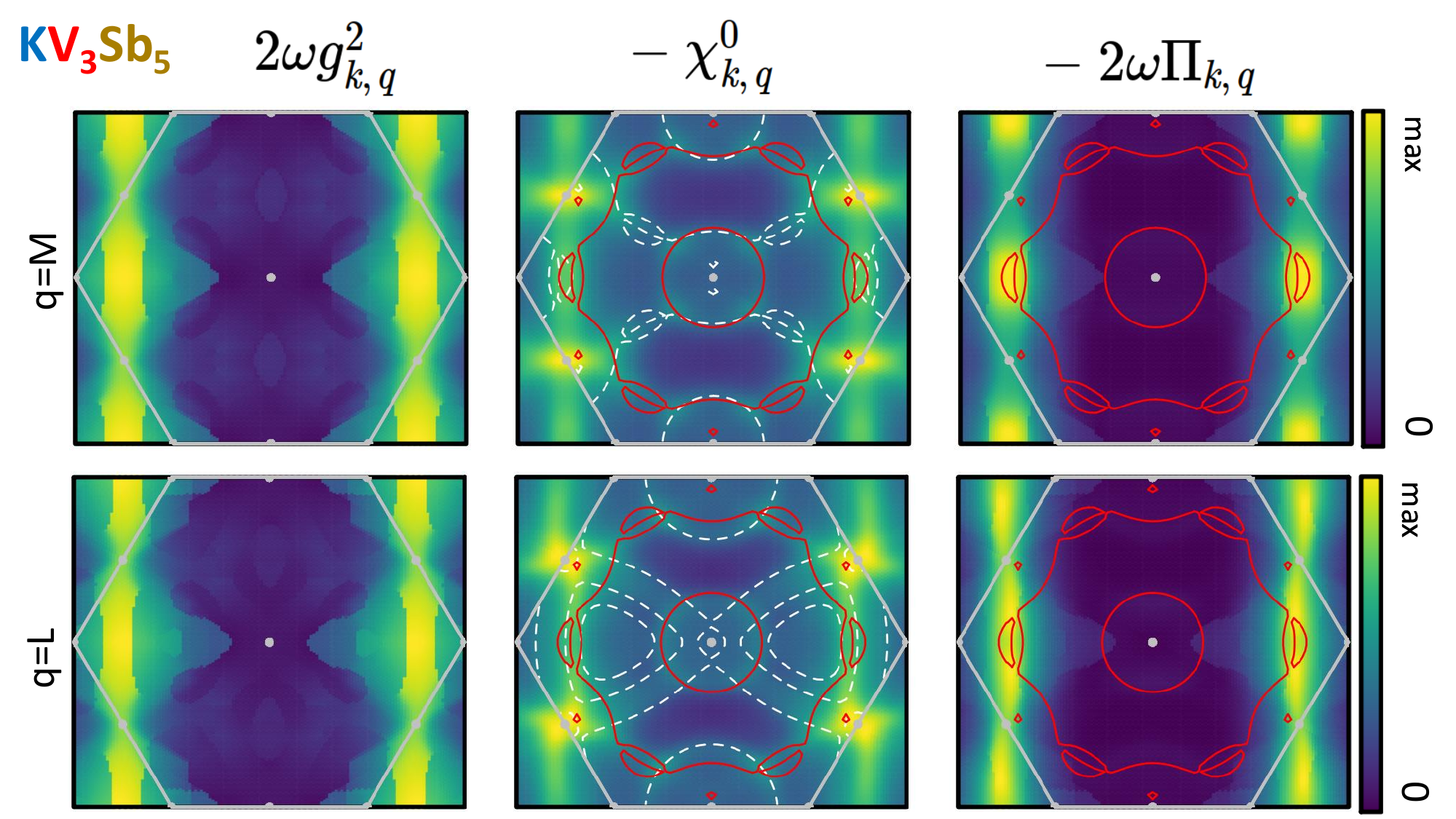}
    \caption{$\vec k$ momentum-resolved phonon fluctuation diagnostics of the instability mode at $\vec q = \mathrm M$ and $\vec q = \mathrm L$ for $k_z = 0$. The coupling matrix elements $2\omega g^2$ (right), the electronic susceptibility $\chi^0$ (middle) and the phonon self-energy $2 \omega \varPi$ (left) of KV\s3Sb\s5 are shown. The solid red lines indicate the Fermi surface and the dashed lines Fermi surface shifted by $q = \mathrm L$ inside the electronic \ac{BZ}.}

    \label{fig:K_fluc}
\end{figure*}

\begin{figure*}
 \includegraphics[width=0.95\linewidth,trim={0cm 0cm 0cm 0cm},clip]{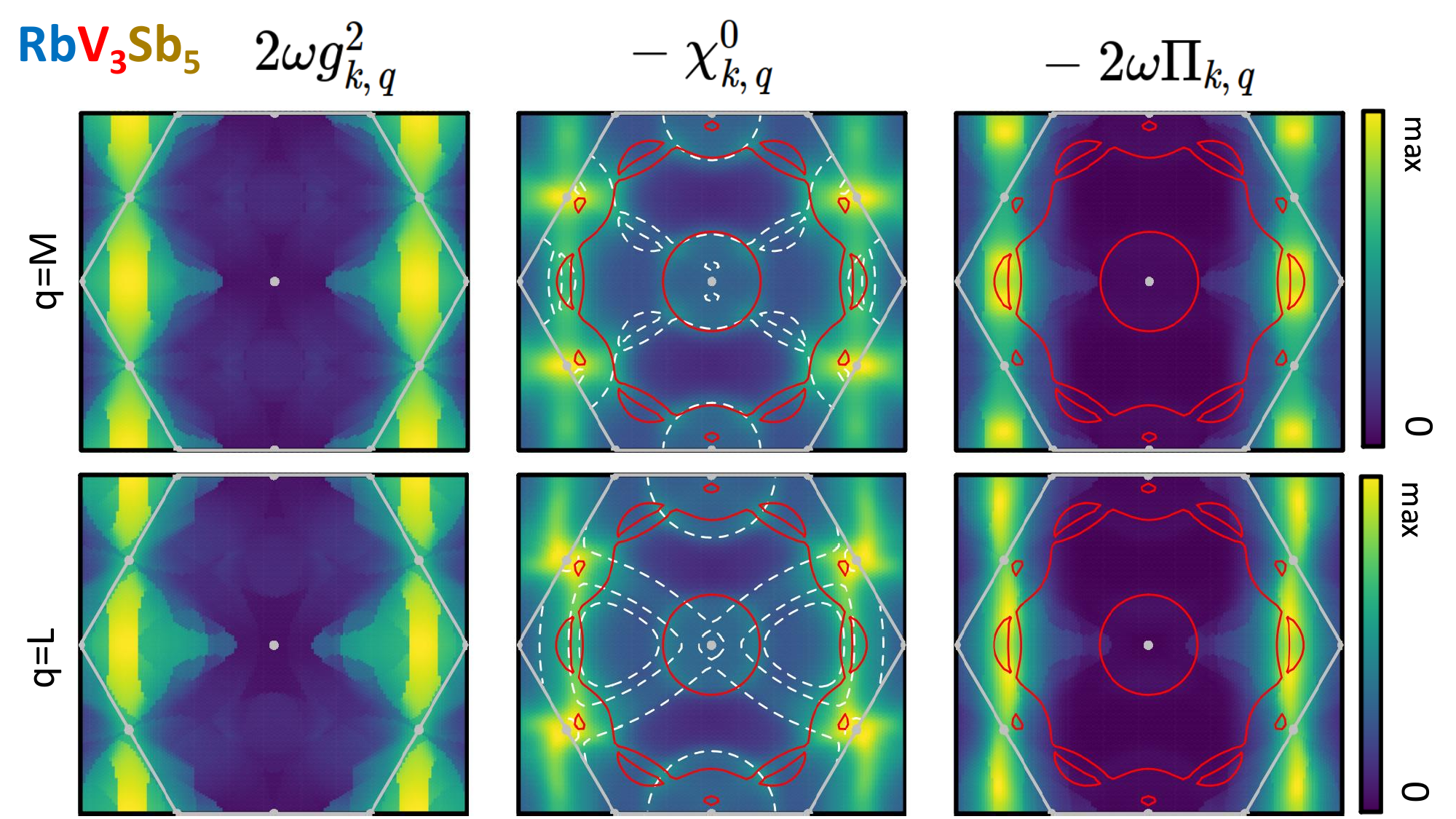}
    \caption{$\vec k$ momentum-resolved phonon fluctuation diagnostics of the instability mode at $\vec q = \mathrm M$ and $\vec q = \mathrm L$ for $k_z=0$. The coupling matrix elements $2\omega g^2$ (right), the electronic susceptibility $\chi^0$ (middle) and the phonon self-energy $2 \omega \varPi$ (left) of RbV\s3Sb\s5 are shown. The solid red lines indicate the Fermi surface and the dashed lines Fermi surface shifted by $q = \mathrm L$ inside the electronic \ac{BZ}.}
    \label{fig:Rb_fluc}
\end{figure*}

\begin{figure*}
 \includegraphics[width=0.95\linewidth,trim={0cm 0cm 0cm 0cm},clip]{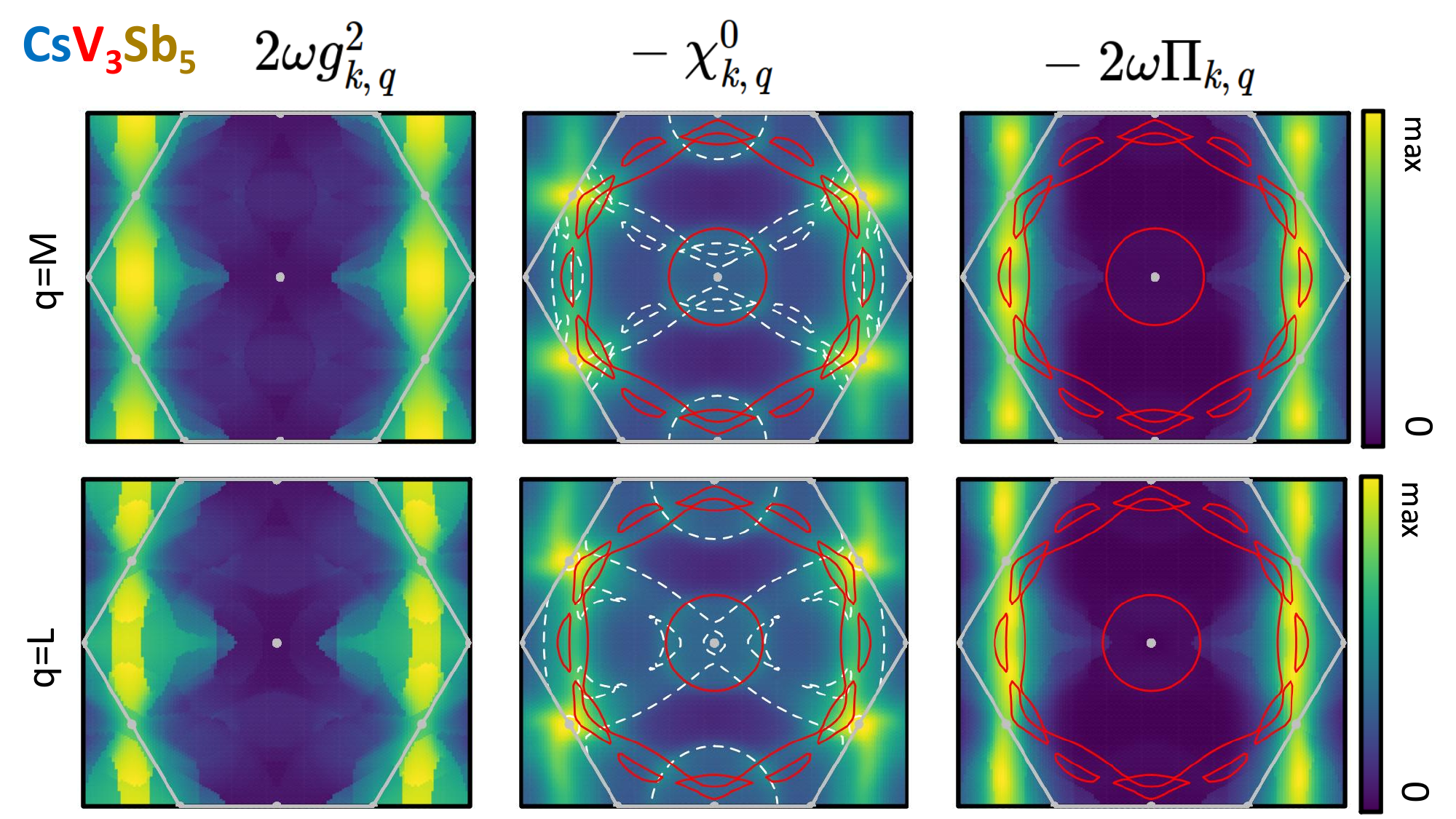}
    \caption{$\vec k$ momentum-resolved phonon fluctuation diagnostics of the instability mode at $\vec q = \mathrm M$ and $\vec q = \mathrm L$ for $k_z=0$. The coupling matrix elements $2\omega g^2$ (right), the electronic susceptibility $\chi^0$ (middle) and the phonon self-energy $2 \omega \varPi$ (left) of CsV\s3Sb\s5 are shown. The solid red lines indicate the Fermi surface and the dashed lines Fermi surface shifted by $q = \mathrm L$ inside the electronic \ac{BZ}.
    }
    \label{fig:Cs_fluc}
\end{figure*}

\begin{figure*}
 \includegraphics[width=0.95\linewidth,trim={0cm 0cm 0cm 0cm},clip]{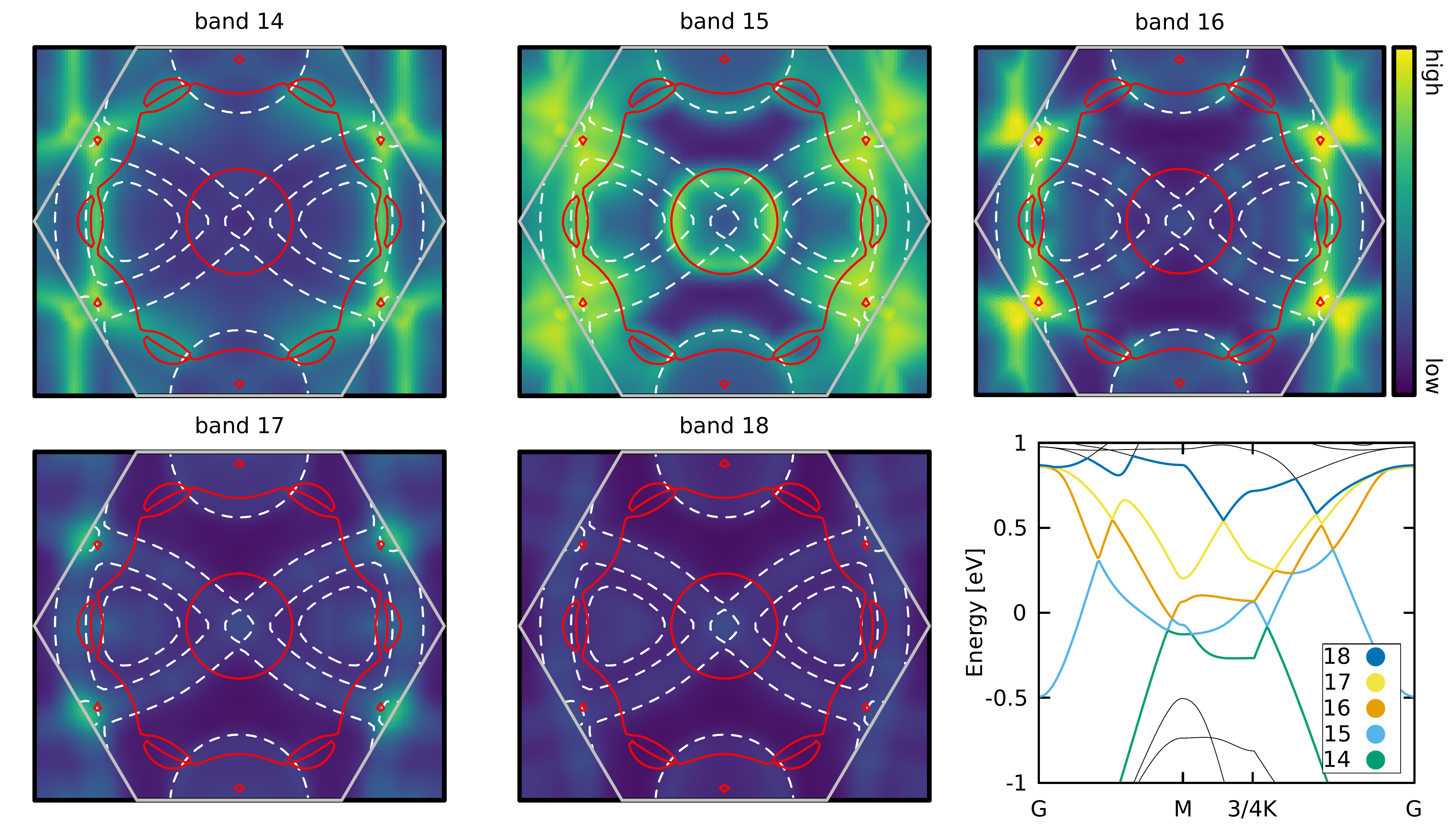}
    \caption{$\vec k$ momentum-resolved electronic susceptibility of KV\s3Sb\s5 for each band around the Fermi level for $\vec q = \mathrm L$ and $k_z=0$. The highlighted bands are considered for the electron-phonon coupling. The solid red lines indicate the Fermi surface and the dashed lines Fermi surface shifted by $q = \mathrm L$ inside the electronic \ac{BZ}.}
    \label{fig:sus_res}
\end{figure*}

\begin{figure*}
 \includegraphics[width=0.95\linewidth,trim={0cm 0cm 1.75cm 0.5cm},clip]{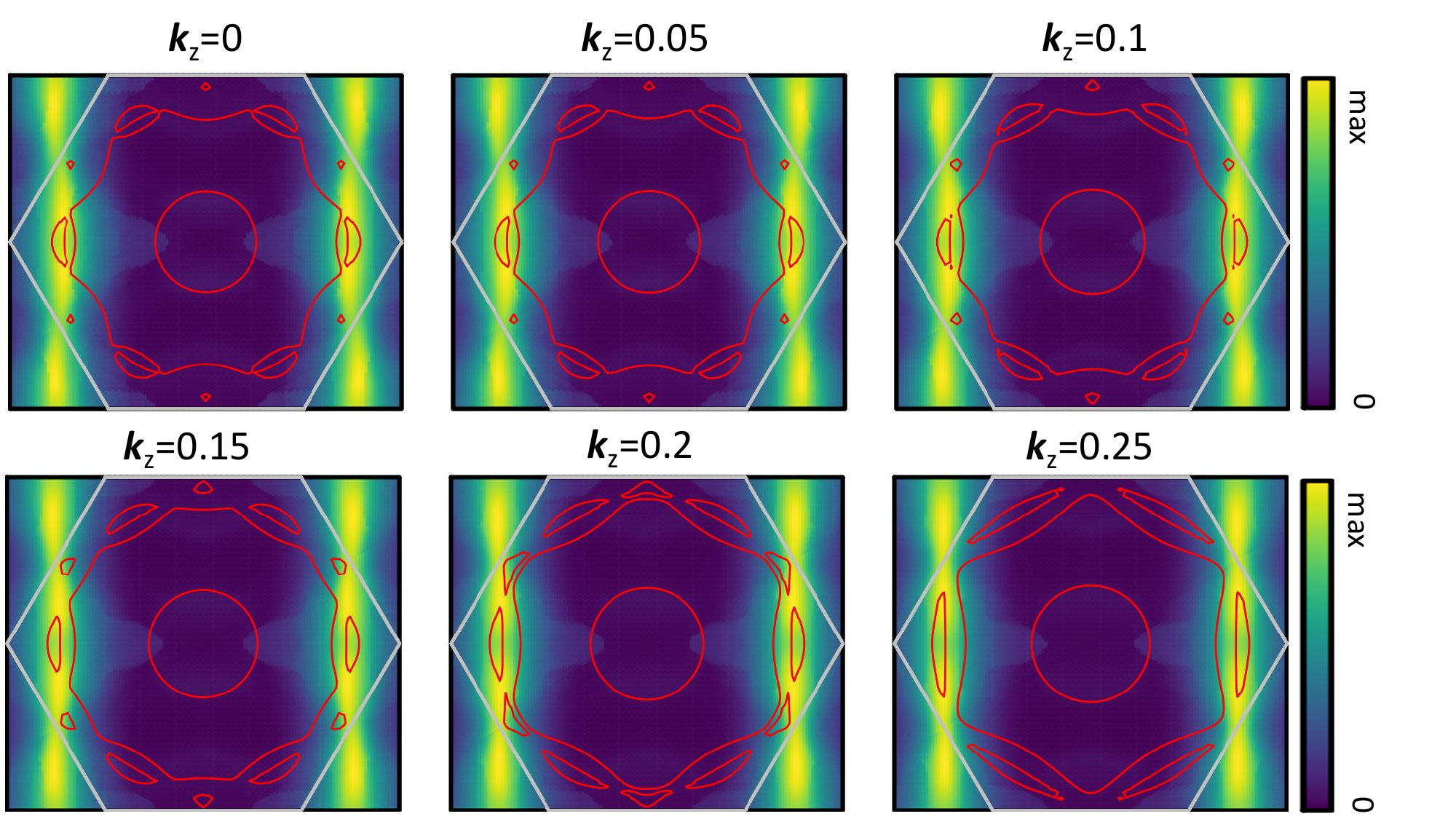}
    \caption{$\vec k$ momentum-resolved phonon self-energy of KV\s3Sb\s5 for $\vec q = \mathrm L$ and different $k_\text{z}$ values. The solid red lines indicate the Fermi surface inside the electronic \ac{BZ}.}
    \label{fig:kz_self}
\end{figure*}

\begin{figure*}
 \includegraphics[width=0.95\linewidth,trim={0cm 0cm 1.75cm 0.5cm},clip]{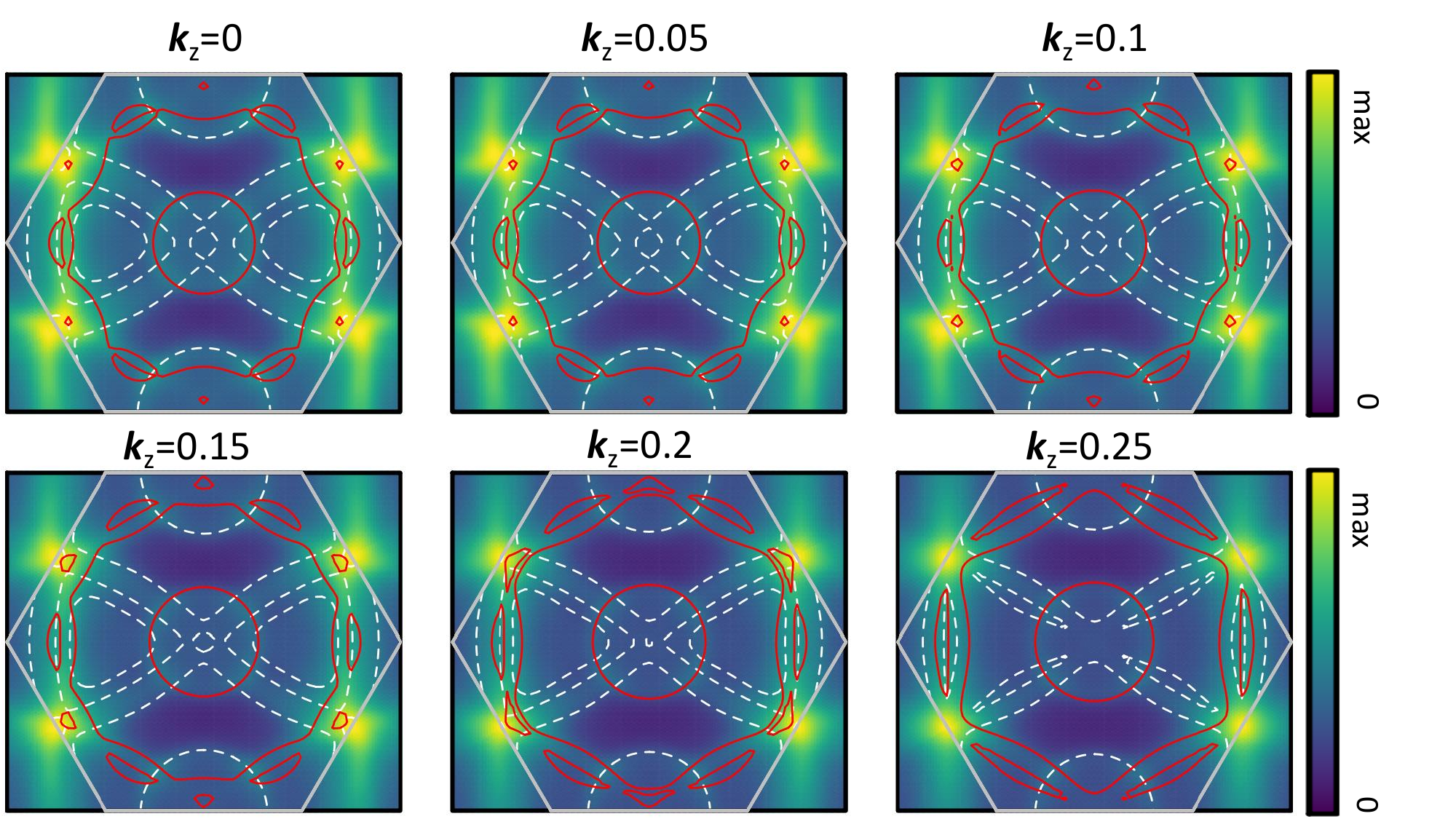}
    \caption{$\vec k$ momentum-resolved electronic susceptibility of KV\s3Sb\s5 for $\vec q = \mathrm L$ and different $k_\text{z}$ values. The solid red lines indicate the Fermi surface and the dashed lines Fermi surface shifted by $q = \mathrm L$ inside the electronic \ac{BZ}.}
    \label{fig:kz_sus}
\end{figure*}

\begin{figure*}
 \includegraphics[width=0.95\linewidth,trim={0cm 0cm 0cm 0cm},clip]{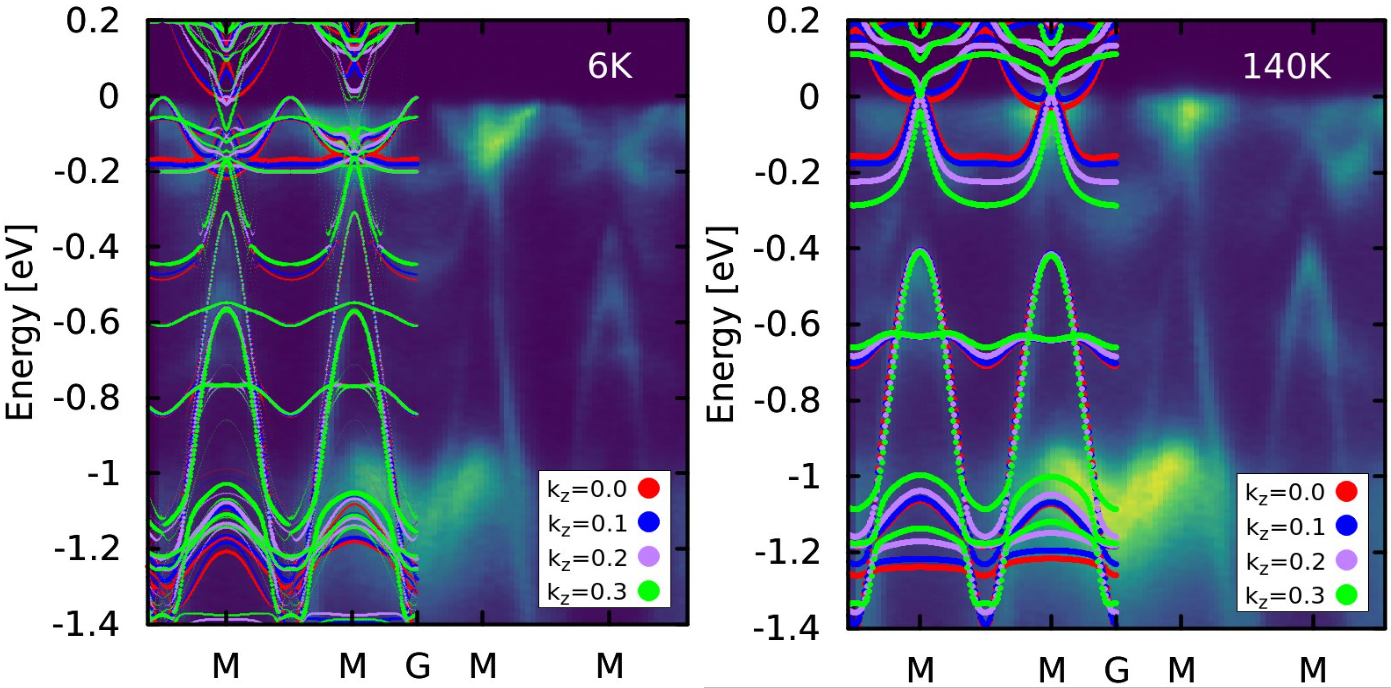}
    \caption{\ac{ARPES} band structure below and above the \ac{CDW} temperature overlayed with \ac{DFT} bands of the primitive cell and the unfolded \ac{DFT} band structure of staggered \ac{ISD} $2 \times 2 \times 2$ phase for different $k_z$.}
    \label{fig:kz_bands}
\end{figure*}


\bibliography{biblio}